\documentstyle[12pt,preprint,aps,floats]{revtex}

\tighten

\input epsf.tex

\def\AP#1#2#3{{Ann. Phys. } {\bf #1}, #2 (#3)}

\def\AP#1#2#3{Astropart. Phys. {\bf #1}, #2 (#3)}

\def\MPLA#1#2#3{Mod. Phys. Lett. {\bf A#1}, #2 (#3)}
\def\NPB#1#2#3{Nucl. Phys. {\bf B#1}, #2 (#3)}

\def\PLB#1#2#3{Phys. Lett. {\bf B#1}, #2 (#3)}
\def\PLBold#1#2#3{Phys. Lett. {\bf#1B}, #2 (#3)}
\def\PRep#1#2#3{Phys. Rep. {\bf#1}, #2 (#3)}
\def\PRD#1#2#3{Phys. Rev. {\bf D#1}, #2 (#3)}
\def\PRL#1#2#3{Phys. Rev. Lett. {\bf#1}, #2 (#3)}

\newcommand{\postscript}[2]{\setlength{\epsfxsize}{#2\hsize}
   \centerline{\epsfbox{#1}}}
\newcommand{\sign}{\:\!\text{sgn}\:\!}
\newcommand{\mgaugino}{M_{1/2}}

\newcommand{\mgut}{M_{\text{GUT}}}

\newcommand{\tb}{\tan\beta}

\newcommand{\gev}{\text{ GeV}}
\newcommand{\tev}{\text{ TeV}}
\newcommand{\Omegachi}{\Omega_{\chi}}
\newcommand{\be}{\begin{equation}}
\newcommand{\ee}{\end{equation}}

\begin{document}

\draft

\renewcommand{\thefootnote}{\fnsymbol{footnote}}
\setcounter{footnote}{0}

\preprint{
\noindent
\hfill
\begin{minipage}[t]{3in}
\begin{flushright}
IASSNS--HEP--00--24\\
FERMILAB--PUB--00/074--T\\
hep-ph/0004043\\
3 May 2000\\
\end{flushright}
\end{minipage}
}

\title{
\vskip 0.5in
Neutralino Dark Matter in Focus Point Supersymmetry}

\author{
Jonathan L.~Feng$^a$, Konstantin T.~Matchev$^b$, and 
Frank Wilczek$^a$
\vskip 0.2in
}

\address{
  ${}^{a}$
  School of Natural Sciences, Institute for Advanced Study\\
  Princeton, NJ 08540, U.S.A. 
\vskip 0.1in
  ${}^{b}$
  Theoretical Physics Department,
  Fermi National Accelerator Laboratory\\
  Batavia, IL 60510, U.S.A.}

\maketitle

\begin{abstract}

In recent work, it has been argued that multi-TeV masses for scalar
superpartners are not unnatural. Indeed, they appear to have
significant phenomenological virtues.  Here we explore the
implications of such `focus point' supersymmetry for the dark matter
problem.  We find that constraints on relic densities do not place
upper bounds on neutralino or scalar masses. We demonstrate that, in
the specific context of minimal supergravity, a cosmologically stable
mixed gaugino-Higgsino state emerges as an excellent, robust dark
matter candidate.  We estimate that, over a wide range of the unknown
parameters, the spin-independent proton-neutralino cross sections fall
in the range accessible to planned search experiments.

\end{abstract}

\vspace{1.3cm}
\centerline{\sl To appear in Phys.~Lett.~B}


\newpage

\renewcommand{\thefootnote}{\arabic{footnote}}
\setcounter{footnote}{0}

\section{Introduction}
\label{sec:introduction}

The idea that supersymmetry is an approximate symmetry of Nature,
broken at a `low' scale, is best and most concretely motivated by two
general arguments.  First, it addresses the question of why radiative
corrections to the electroweak symmetry breaking scale, due to the
exchange of virtual particles at some much larger unification or
Planck scale, do not pull the electroweak scale to a much higher value
(the gauge hierarchy problem~\cite{SM}).  Second, it brings the
unification of couplings calculation, which does not quite work if we
use only the particles of the Standard Model, into adequate agreement
with experiment~\cite{DRW}.  However neither of these arguments, on
the face of it, yields very specific constraints on the spectrum of
masses of the supersymmetric ($R$-parity odd) partners of ordinary
($R$-parity even) particles: the first because it is inherently
semi-quantitative, the second because it depends on these masses only
logarithmically.
 
On the other hand, it is becoming notorious that the existence of
superpartners with masses below 1 TeV is difficult to reconcile with
established limits on various flavor violating effects, unless one
assumes accurate degeneracy of squarks and of sleptons.  Even if one
assumes such degeneracy, supersymmetric contributions to $CP$
violation and (in unified theories) proton decay become uncomfortably
large, unless the squarks and sleptons are heavy.

Therefore one is motivated to examine supersymmetric models with large
scalar superpartner masses.  The issue immediately arises: To what
extent can such models, when critically scrutinized, embody the motivating
features we mentioned in our opening paragraph?

Recently it has been argued, using a seemingly reasonable objective
definition of naturalness, that all squark and slepton masses can be
taken well above 1 TeV with no loss of
naturalness~\cite{FMMshort,FMMlong}.  The mathematical basis of this
result is the existence of focus points in renormalization group
trajectories, which render the weak scale ({\em i.e.}, the Higgs
potential) largely insensitive to variations in unknown supersymmetry
parameters~\cite{FMMshort,FMMlong,FMMproc}.  We should note that while
in these models the squark and slepton masses are unusually large
(compared to conventional wisdom), the electroweak gaugino and
Higgsino particle masses are generically well below 1 TeV.

In the focus point models, moreover, the quantitative aspect of the
unification of coupling constants is actually slightly improved
relative to more traditional supersymmetry models, reducing the need
for significant high scale threshold corrections~\cite{FMinprep}.

Besides preserving and sharpening the original motivations for
supersymmetry, the focus point models with heavy scalars ameliorate
the problems noted above~\cite{FMinprep}.  Specifically, constraints
on $CP$ violation from electric dipole moments can be satisfied even
with ${\cal O} (1)$ phases in the fundamental parameters, and the
predicted rate for nucleon decay --- being proportional to the square
of gaugino and inversely proportional to the fourth power of
squark/slepton masses --- is suppressed and less dangerous, though
still perhaps accessible.

In this note we will consider the implications of focus point
supersymmetry for neutralino dark matter.  We will concentrate on the
focus point mechanism in its simplest `default' incarnation, assuming
a universal scalar mass within minimal supergravity. The focus point
mechanism can also be realized in alternative models featuring
gravity-~\cite{FMMshort,FMMlong}, gauge-~\cite{A}, and
anomaly-mediated~\cite{FM} supersymmetry breaking.  In addition, the
requirement of a universal scalar mass, though sufficient, is not
necessary.  For the precise set of requirements, see
Ref.~\cite{FMMlong}.

Of course, the cosmology of minimal supergravity (with universal
scalar mass $m_0 \leq 1 \tev$) has been studied extensively.
Prominent among conclusions one finds in the literature are the
following:

\bigskip

\noindent (i) In essentially all of parameter space, the lightest
neutralino $\chi$ is Bino-like, with gaugino fraction $R_{\chi} >
0.9$~\cite{Snowmass}. (See below for the definition of $R_{\chi}$.)

\noindent (ii) The neutralino relic abundance $\Omegachi h^2$
increases as $m_0$ increases.  The requirement that neutralinos not
overclose the universe, along with the requirement that the lightest
supersymmetric particle (LSP) be neutral, then typically leads to
stringent upper bounds on scalar and/or gaugino masses, independent of
naturalness considerations~\cite{OS,GKT,LYN,KKRW,BB,AN3,EFO,EFOS,AC,EGO}.

\noindent (iii) Dark matter detection rates decrease as $m_0$
increases.  The rates predicted range widely over parameter space, but
the prospects for discovery at ongoing experiments grow increasingly
dim as $m_0$ increases~\cite{ANdetection}.

\bigskip

In our analysis, we confirm these conclusions --- for $m_0 \leq 1
\tev$.  However, we find that they fail for focus point models with
$m_0 > 1 \tev$.  Instead, we find that in such models:

\bigskip

\noindent (i) The lightest neutralino is a gaugino-Higgsino mixture
over much of parameter space.

\noindent (ii) Neutralino relic abundances do not overclose the
universe.  Rather, they tend to lie in the cosmologically interesting
range.  Therefore, in particular, cosmological considerations alone do
not place stringent upper limits on superpartner masses, even apart
from possible conspiracies involving poles~\cite{DNOmega,DNRY} or
co-annihilation~\cite{EFO,Dreesstop}.

\noindent (iii) Direct detection rates may be fairly large, with
proton-neutralino cross sections plausibly falling within the range
$10^{-6}$-$10^{-8}$ pb.  (See the more detailed discussion and plots
below.)  

\bigskip

These results are especially interesting since ongoing and
planned dark matter detection experiments promise to cover the
indicated 
range.

\section{Neutralinos in Focus Point Supersymmetry}
\label{sec:neutralinos}

The models we consider, like many supersymmetric models, contain a
natural cold dark matter candidate~\cite{G}.  Assuming that $R \equiv
(-)^{B+L+2S}$ is accurately conserved, the lightest supersymmetric
($R$-odd) particle is stable.  We will be considering models where,
with no special adjustment, this is a neutral weakly interacting
massive particle (WIMP), which is commonly called the neutralino.

The relic densities and detection rates of neutralino dark matter can
be calculated, assuming straightforward extrapolation of Big Bang
cosmology to $T\sim 10 \gev$~\cite{LW}.  In principle they depend on
details of the entire supersymmetric spectrum, but in practice,
especially in focus point models, they are mainly determined by the
properties of the neutralino LSP itself.  

Assuming $R$-parity conservation and minimal field content, at tree
level the neutralino mass matrix is
\begin{equation}
\label{neumass}
\left( \begin{array}{cccc}
M_1        &0       
&-m_Z \cos\beta\, \sin \theta_W & m_Z \sin\beta\, \sin \theta_W \\
0          &M_2     & m_Z \cos\beta\, \cos
\theta_W &-m_Z \sin\beta\, \cos
\theta_W \\
-m_Z \cos\beta\, \sin \theta_W  & m_Z \cos\beta\, \cos
\theta_W &0       &-\mu      \\
 m_Z \sin\beta\, \sin \theta_W  &-m_Z \sin\beta\, \cos
\theta_W &-\mu    &0 \end{array}
\right)
\end{equation}
in the basis $(-i\tilde{B},-i\tilde{W}^3, \tilde{H}^0_u,
\tilde{H}^0_d)$.  Here $M_1$ and $M_2$ are the soft Bino and Wino
masses, $\mu$ is the Higgsino mass parameter, and $\tb = \langle H^0_u
\rangle / \langle H^0_d \rangle$ is the ratio of Higgs vacuum
expectation values.  We parametrize the gaugino/Higgsino content of
the lightest neutralino according to
\begin{equation}
\chi = a_1 (-i \tilde{B}) + a_2 (-i \tilde{W}^3) + a_3 \tilde{H}^0_u
+ a_4 \tilde{H}^0_d \ ,
\end{equation}
and define the gaugino fraction of $\chi$ to be 
\begin{equation}
R_{\chi} \equiv |a_1|^2 + |a_2|^2 \ .
\end{equation}

In the framework of minimal supergravity, there are 4 continuous
parameters and 1
binary choice:
\begin{equation} 
m_0, \mgaugino, A_0, \tb, \sign (\mu) \ .  
\end{equation} 
Here, $m_0$, $\mgaugino$, and $A_0$ are, respectively, the universal
scalar mass, gaugino mass, and trilinear scalar coupling at the grand
unified theory (GUT) scale $\mgut \simeq 2 \times 10^{16} \gev$.
Given values for these input parameters, all the couplings and masses
of the weak scale Lagrangian are determined through renormalization
group (RG) evolution.  In our work we use two-loop RG
equations~\cite{2loop RGEs} and include one-loop threshold corrections
from supersymmetric particles to the gauge and Yukawa coupling
constants~\cite{BMP,PBMZ}. We minimize the Higgs potential after
including all one-loop effects, and include one-loop corrections in
all superpartner masses~\cite{PBMZ}.  This framework imposes several
specific relations among the weak scale supersymmetry parameters.  In
particular, $M_1 \simeq M_2 / 2 \simeq 0.4 \mgaugino$, and $|\mu|$ is
fixed by the condition of electroweak symmetry breaking, which, at
tree-level, is
\begin{equation}
\frac{1}{2} m_Z^2 = \frac{m_{H_d}^2 - m_{H_u}^2 \tan^2\beta }
{\tan^2\beta -1} - \mu^2 \approx - m_{H_u}^2 - \mu^2 \ ,
\label{mZ}
\end{equation}
where the last relation is valid for moderate and large $\tb$.

The essence of the focus point mechanism is the observation that the
weak scale value of $m_{H_u}^2$ is remarkably insensitive to
variations in the fundamental GUT scale supersymmetry parameters, even
for multi-TeV $m_0$.  By Eq.~(\ref{mZ}), therefore, for $\tb \agt
5$~\cite{FMMlong}, the electroweak scale is insensitive to variations
in these parameters, and in this sense multi-TeV values of $m_0$ are
natural.

\begin{figure}[t]
\postscript{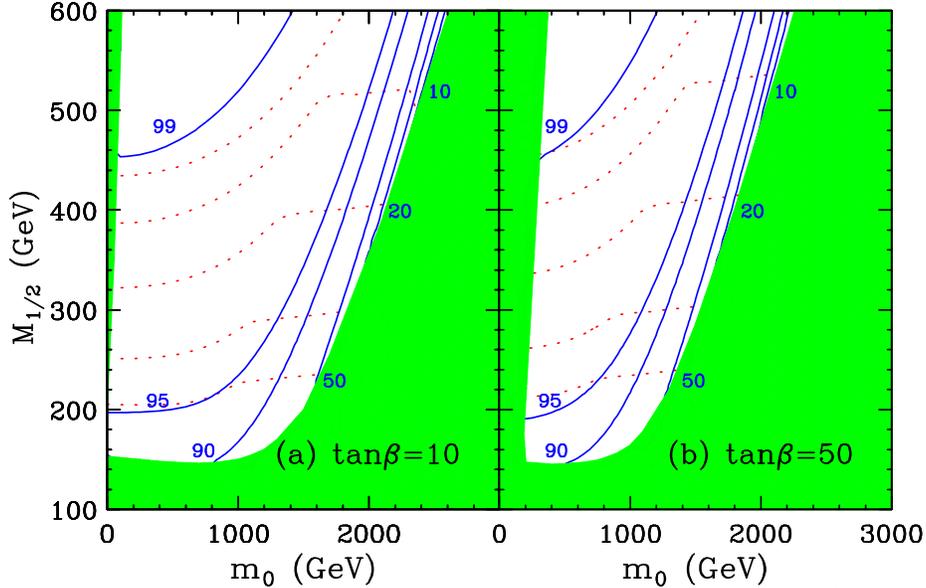}{0.74}
\caption{Contours of constant LSP gaugino fraction $R_{\chi}$ (in
percent) in the $(m_0, \mgaugino)$ plane for $A_0=0$, $\mu>0$, and two
representative values of $\tb$. The shaded regions are excluded by the
requirement that the LSP be neutral (top left) and by the chargino
mass limit of 95 GeV (bottom and right).  Dashed contours are for the
fine tuning parameter $c = 20$, 30, 50, 75 and 100, from below (see
text).}
\label{fig:gfraction}
\end{figure}

Contours of gaugino fraction $R_{\chi}$ are presented in
Fig.~\ref{fig:gfraction}.  (For reference, contours for the
fine-tuning parameter $c$, as defined in Ref.~\cite{FMMshort}, are
also given in Fig.~\ref{fig:gfraction}. While large $M_{1/2}$ leads to
large fine-tuning, large $m_0$ does not, as a result of the focus
point discussed above.)  For $m_0 \leq 1 \tev$, the lightest
neutralino is nearly pure Bino, with $R_{\chi} \agt 0.9$.  This
well-known result arises from the circumstance that RG evolution
typically drives $m_{H_u}^2$ large (relative to the gaugino masses)
and negative, which by Eq.~(\ref{mZ}) implies $|\mu|$ much larger than
$M_1$ and $M_2$.

For $m_0 > 1 \tev$, however, $\chi$ contains significant amounts of
Higgsino over much of the parameter space.  For large and moderate
values of $\tb$, large values of $m_0$ generate positive corrections
to $m_{H_u}^2$, rendering it less negative.  Thus if all other
parameters are fixed, we find that as $m_0$ increases, $|\mu|$
decreases, and eventually $|\mu| \sim M_1, M_2$, which leads to
significant mixing between Higgsino and gaugino states.  As $m_0$
increases further, the Higgsino content of $\chi$ increases until
ultimately one enters the domain $|\mu| < 95\gev$, which is excluded
by limits from chargino searches at LEP II~\cite{charginolimit}.  At
that point, we have entered the shaded region of
Fig.~\ref{fig:gfraction}.

Note that the above discussion holds for $\tb \agt 5$ where the focus
point mechanism is operative.  For small $\tb$, $\mu$ becomes
sensitive to $m_{H_d}^2$, and large $m_0$ gives {\em negative}
contributions to $m_{H_u}^2$.  Both of these effects imply that, as
$m_0$ increases, $|\mu|$ also increases, and so there is no mixed
gaugino-Higgsino region for small $\tb$.


\section{Relic Abundance}
\label{sec:omega}

There is ample and increasingly precise evidence that the energy
density of the universe is not dominated by the observed luminous
matter~\cite{DMreviews}.  The evidence is conveniently expressed in
terms of ratios $\Omega_i \equiv \rho_i / \rho_c$, where the $\rho_i$
are energy densities and $\rho_c$ is the critical density, and $h
\approx 0.65 \pm 0.1$, the Hubble parameter in units of 100 km
s$^{-1}$ Mpc$^{-1}$.  The luminous matter density is roughly
$\Omega_{\text{lum}} \sim 0.005$, and the successful predictions of
Big Bang nucleosynthesis require baryon density $\Omega_b h^2 \approx
0.02$.  At the same time, rotation curves of spiral galaxies require
matter density $\Omega_m \agt 0.1$, and a variety of observations,
ranging from the velocities of galaxies within galactic clusters to
the luminosities of Type Ia supernovae, favor values in the range $0.2
\alt \Omega_m \alt 0.6$.  There is therefore a need for both baryonic
and non-baryonic dark matter.  For non-baryonic dark matter, which is
of interest here, galactic rotation curves require $\Omega_{\text{DM}}
h^2 \agt 0.025$, and large scale measurements suggest $0.1 \alt
\Omega_{\text{DM}} h^2 \alt 0.3$.

Neutralinos remain in thermal equilibrium until their pair
annihilation rate drops below the Hubble expansion rate.  The relic
density of neutralinos is therefore determined primarily by their
cross section for pair annihilation in the early universe.
Neutralinos annihilate to fermion pairs through the $t$-channel
exchange of sfermions $\tilde{f}$, and via $\chi \chi \to Z, A, h, H
\to f \bar{f}$, where $A$ is the $CP$-odd Higgs boson, and $h$ and $H$
are the lighter and heavier $CP$-even Higgs bosons. When kinematically
accessible, annihilation to two body states of gauge and Higgs bosons
are also relevant.

The calculation of neutralino relic density has been discussed at
length in the literature.  We use the results of Drees and
Nojiri~\cite{DNOmega,DNsigma}, as implemented by Jungman,
Kamionkowski, and Griest~\cite{JKG}.  $S$- and $P$-wave contributions
to all tree-level processes with two-body final states are included.

\begin{figure}[t]
\postscript{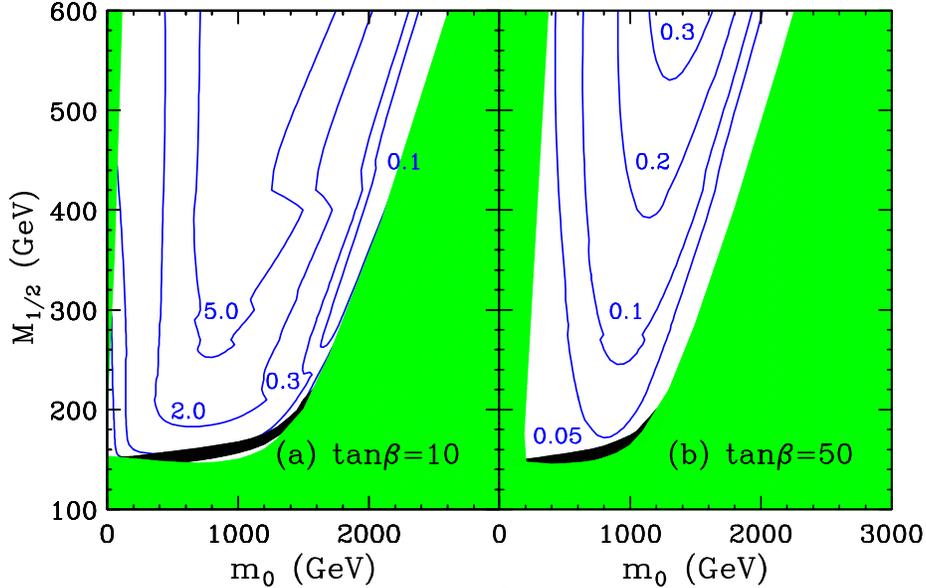}{0.74}
\caption{Contours of constant relic density $\Omegachi h^2$ in the
$(m_0, \mgaugino)$ plane for $A_0=0$, $\mu>0$, and two representative
values of $\tb$. In the black shaded region, $|2 m_{\chi} - m_h| <
5\gev$, and the $h$ pole becomes important.  The light shaded regions
are as in Fig.~\ref{fig:gfraction}.}
\label{fig:omega}
\end{figure}

In Fig.~\ref{fig:omega} we display the neutralino relic density
$\Omegachi h^2$ in the $(m_0, \mgaugino)$ plane for two representative
values of $\tb$, $A_0=0$, and $\mu>0$.  These plots exhibit the most
important parameter dependencies. Variations in $A_0$ within its
natural range have little effect, as the $A$ parameter flows to a
fixed point at the weak scale.  Similarly, changing the sign of $\mu$
does not produce qualitatively new results.  We have chosen the sign
less constrained by $b\to s \gamma$ (see below).

In most of parameter space, the expected accuracy in
Fig.~\ref{fig:omega} is ${\cal O} (10 \%)$~\cite{JKG}. More
significant errors are possible in special regions.  First, a more
refined treatment is necessary just below annihilation
thresholds~\cite{GG,GS}.  Such a treatment will smooth out the kinks
in the contours of Fig.~\ref{fig:omega}, which are caused by the
opening of annihilation channels such as $WW$ and $ZZ$.  Second,
co-annihilation is important very near the left and right borders of
the allowed region, where the LSP is nearly degenerate with
staus~\cite{EFO,Gomez} and Higgsinos~\cite{MY}, respectively.  Finally,
$s$-channel poles also require a more careful
analysis~\cite{GG,GS,ANpoles,BB}.  The $Z$ pole possibility, $2
m_{\chi} = m_Z$, is now essentially eliminated by the chargino mass
bound of 95 GeV~\cite{charginolimit}, but a $h$ pole is possible near
the bottom of the allowed region, as indicated in
Fig.~\ref{fig:omega}.  $H$ and $A$ poles, though possibly significant
for high $\tb$ in the $m_0 \leq 1 \tev$ region~\cite{DNRY,BK}, are
absent for $m_0 > 1 \tev$ for all $\tb \alt 50$ and $\mu >0$.  (The
sign of $\mu$ enters through the finite supersymmetric threshold
corrections to the bottom Yukawa coupling~\cite{ybcorr}.)  Thus,
although special effects warrant more sophisticated analysis in
limited regions of parameter space, they are absent in the bulk of the
focus point region.  They do not affect our main conclusions.

For sub-TeV $m_0$, we see that $\Omegachi h^2$ is a monotonically
increasing function of $m_0$.  In this region, the LSP is nearly
Bino-like, and so the dominant process is $\chi \chi \to f \bar{f}$
through $t$-channel $\tilde{f}$ exchange.  As $m_0$ and
$m_{\tilde{f}}$ increase, this annihilation process is suppressed, and
the relic density grows.  This behavior, along with the requirement of
a neutral LSP, has led many authors to conclude that upper bounds on
the relic density typically impose stringent upper limits on
superpartner masses~\cite{OS,GKT,LYN,KKRW,BB,AN3,EFO,EFOS}. As these
results are independent of, and less subjective than, considerations
based on naturalness or the fine structure of gauge coupling
unification, they have often been taken, by the authors themselves or
others, as robust phenomenological upper bounds.  If correct, this
conclusion would have important implications for supersymmetry
searches at the Tevatron and the LHC, and in planning for future
linear colliders~\cite{BB,AC,EGO}.

But when we consider $m_0 > 1 \tev$, motivated by the focus point or
otherwise, we find that the behavior of this function reverses.  The
reason for this behavior is not difficult to locate.  For although as
$m_0$ increases the $t$-channel sfermion exchange process is more and
more suppressed, as noted in Sec.~\ref{sec:neutralinos}, at the same
time the LSP gradually acquires a significant Higgsino component.
Because of this, other diagrams become unsuppressed, and pair
annihilation again becomes efficient.  The cosmologically interesting
region, with a mixed gaugino-Higgsino LSP and $0.1 \alt \Omegachi h^2
\alt 0.3$, extends at least to $m_0 \sim 10 \tev$, $M_{1/2} \sim 6
\tev$ and $m_{\chi} \sim 2.5 \tev$.  Thus, for all collider
applications, cosmological considerations do {\em not} provide upper
bounds on superpartner masses, even in the constrained framework of
minimal supergravity.  It has been noted previously that cosmological
bounds might be weakened or eliminated by the effects of
co-annihilation~\cite{EFO,Dreesstop} or poles~\cite{DNOmega,DNRY}.
Here, we see that even without appealing to such conspiratorial
effects, sweeping claims that cosmology provides stringent upper
bounds on superpartner masses are unfounded. Note that the
cosmologically interesting region of parameter space with $m_0 >
1\tev$ is in no sense small.  For fixed $M_{1/2}$, this region extends
over one to several hundred GeV in $m_0$, depending on $\tb$.  It is
comparable in area to the allowed region with $m_0 \leq 1 \tev$.

For all $\tb \agt 5$, we find relic densities in the cosmologically
interesting range $0.025 < \Omegachi h^2 < 0.3$ over most of the focus
point regime with $m_0 > 1.5 \tev$.  In fact, as can be seen from
Fig.~\ref{fig:omega}, for $\tb = 50$, $\Omegachi h^2$ never saturates
the upper bound, even for $m_0 \sim 1 \tev$.  This additional
suppression of $\Omegachi h^2$ is not due to any enhancement of the
LSP's Higgsino component from large $\tb$.  As can be seen in
Fig.~\ref{fig:gfraction}, the gaugino purity of the LSP is largely
independent of $\tb$.  Rather it results from two enhancements to the
process $\chi \chi \to A \to f \bar{f}$.  First, the $Af\bar{f}$
coupling is proportional to $\tb$ for isospin $-\frac{1}{2}$ fermions.
Second, for large $\tb$, the bottom Yukawa coupling $h_b$ becomes
significant and lowers $m_A$. This can be seen by noting that, for
$h_t = h_b$, symmetry under interchange $t \leftrightarrow b$ implies
that $m_{H_d}^2$ also has a weak scale focus point, and so $m_A \sim
{\cal O}(100\gev) \ll m_0$.

For small $\tb$, as noted in Sec.~\ref{sec:neutralinos}, there is no
mixed gaugino-Higgsino region for large $m_0$.  Thus, $\Omegachi h^2$
grows monotonically as $m_0$ increases.  Low values of $\tb$, such as
$\tb=2$, have been widely considered in the literature.  They are,
however, increasingly disfavored by current null results in Higgs
searches, and, in any case, are less natural than moderate and large
values, as low $\tb$ typically leads to large values of
$|\mu|$~\cite{FMMlong}.

\section{Direct Detection}
\label{sec:direct}

Having shown that focus point models predict cosmologically
interesting densities of dark matter, we now consider their
predictions for dark matter detection rates.  The interactions of
neutralinos with matter are dominated by scalar (spin-independent)
couplings~\cite{DNsigma,JKG}.  These interactions are mediated either
by $t$-channel $h$ and $H$ exchange, or via $\chi q \to \tilde{q} \to
\chi q$.  The former diagrams contain $h\chi\chi$ and $H\chi\chi$
vertices, which are suppressed for Bino-like $\chi$.  They also
contain $h,H$-nucleon vertices.  These are dominated by contributions
arising from fundamental couplings of $h,H$ to the strange quark and
indirectly to gluons through heavy quark loops.

In the calculation of dark matter detection rates, we assume a
Maxwellian $\chi$ velocity distribution with velocity dispersion
$\bar{v} = 270\ \text{km/s}$, and a local dark matter density of
$\rho_0 = 0.3\ \text{GeV/cm$^3$}$.  

A major uncertainty in the estimate of dark matter detection rates
arises from the poorly determined value of the nucleon matrix element
\begin{equation}
\label{fTsDef}
f_{T_s} \equiv \langle N| m_s \bar s s | N \rangle / m_N \ .
\end{equation}
If it is correct to treat the effect of the strange quark mass term
$m_s \bar s s$ in first order perturbation theory, then we can
identify $f_{T_s}$ as the fractional change in the nucleon mass from
what it would be in a world with massless $s$ quarks.  Values as high
as $f_{T_s} = 0.62$ and as low as $f_{T_s} = 0.08$ have been considered
in the literature~\cite{FTS}.  

The traditional strategy for evaluating $f_{T_s}$ is to combine an
`experimental' determination of the $\sigma$-term $\langle N| {m_u +
m_d\over 2} (\bar u u + \bar d d ) |N \rangle$ with a fit of the
flavor $SU(3)$-breaking term $\langle N|{1\over 3} m_s(2 \bar s s -
\bar u u - \bar d d )| N \rangle$ to the baryon octet splitting, and
finally the ratios of light quark masses from the $\pi/K$ mass ratio.
This procedure is problematic, however, in several ways.  First,
theory relates the value of the $\sigma$ term to the value of the $\pi
N$ scattering amplitude at an unphysical point, and a rather elaborate
and numerically unstable extrapolation is necessary (hence the
quotation marks).  Second, in getting to the quantity of interest, the
uncertainty in the $\sigma$ term is amplified by the large factor
$2m_s/(m_u+ m_d)$.  Third, the nucleon matrix element of the $SU(3)$
breaking term is presumably dominated by the valence quark terms, so
that the term of interest is subdominant.  Fourth, and exacerbating
all these, the use of first-order perturbation theory for the baryon
splittings is questionable to begin with.

Clearly it would be very desirable to have a first-principles estimate
of $f_{T_s}$.  Straightforward evaluation in lattice gauge theory
appears very difficult, because three-point amplitudes with
disconnected components are noisy and computationally demanding.  Here
we would like to suggest an alternative strategy that avoids any
measurement of three-point amplitudes.  The basic idea is to vary the
strange quark mass, starting with a very large value and bringing it
down by small increments.  In the heavy quark limit we have a
universal answer by integrating out the heavy quark to obtain a gluon
operator, which is (essentially) the trace of the energy-momentum
tensor with a definite coefficient~\cite{SVZ}.  And if we change the
strange quark mass by a small amount $\Delta m_s$, we have for the
induced change in the nucleon mass
\begin{equation}
\label{msTrick}
\Delta m_N ~=~ \langle N| \Delta m_s \bar s s | N \rangle \ ,
\end{equation}
by first-order perturbation theory.  Of course, it is implicit in this
procedure that one is doing a fully unquenched calculation, with
dynamical quarks (including the strange quark).  Since the error is at
least quadratic in the step size, one should achieve good accuracy by
taking the steps sufficiently small.  Note that, since the nucleon
state is changing at each step of the calculation, it is not correct
to do the whole thing in one jump, despite the apparent linearity of
Eq.~(\ref{msTrick}).  In any case, Eq.~(\ref{msTrick}) allows one to
extract the matrix element of interest by measuring masses only,
without ever involving three-point functions.

No result of this type is currently available, and so at present one
must rely on informed guesswork for a numerical estimate of $f_{T_s}$.
There is a very extensive literature on the subject, which we will not
review here.  We will confine ourselves to two brief observations.

If one treats the strange quark as heavy, rather than light, then one
obtains the universal value $f_{T_{\text{Heavy}}} = 0.074$.  This is
presumably a lower limit.  But of course if one takes the strange
quark mass to zero, then $f_{T_s}$ will likewise vanish.  Thus, in
considering values of $f_{T_s}$ that are many times the heavy quark
value, one is implicitly postulating that this quantity varies very
rapidly as a function of $m_s$, which may be implausible.

Quenched calculations of the nucleon to rho mass ratio, and even more
so calculations with two dynamical quarks, agree remarkably well with
experiment, down to the few percent level.  This fact can only be
consistent with the idea that the strange quark is significantly
perturbing the nucleon mass if the strange quark contributes an
accurately equal proportion of the rho mass.  In general, that would
seem to require a conspiracy.  Just such a ``conspiracy'' does take
place for heavy quarks, since they can be integrated out in favor of a
charge renormalization, whose primary effect is just to change the
overall mass scale (dimensional transmutation).  However, as we have
seen, if it is valid to treat the strange quark as heavy, then its
contribution is small.

While these considerations are far from definitive, they do seem to us
to make the larger end of the range of values discussed for $f_{T_s}$
appear dubious.  In preparing our plots, we have adopted the
conventional value $f_{T_s} = 0.14$. For extreme values in the range
quoted above, the proton-neutralino cross sections plotted may
decrease by roughly a factor of 2, or increase by a factor of
5.
 
\begin{figure}[t]
\postscript{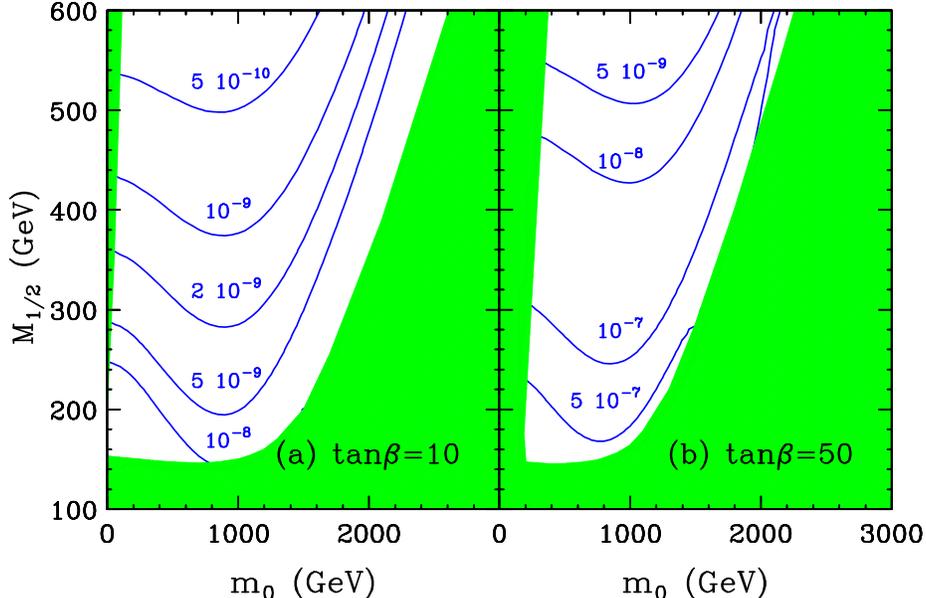}{0.74}
\caption{Contours of proton-$\chi$ cross section $\sigma_P$ in pb in
the $(m_0, \mgaugino)$ plane for $A_0=0$, $\mu>0$, and two
representative values of $\tb$. The shaded regions are as in
Fig.~\ref{fig:gfraction}.}
\label{fig:sigma}
\end{figure}

In Fig.~\ref{fig:sigma}, we present proton-neutralino cross sections
$\sigma_P$.  For $m_0 \alt 1 \tev$, the dominant contribution is
through squarks, and so the interaction rate decreases monotonically
for increasing $m_0$~\cite{ANdetection}.  A naive extrapolation would
then suggest very small detection rates in the focus point region.
However, for $m_0 \agt 1 \tev$, this behavior is reversed --- in this
region, as $m_0$ increases, the LSP's Higgsino content increases, and
so the Higgs boson diagrams become less and less suppressed.  (This
feature was also noticed in Ref.~\cite{MBKM}, where the region of
large $m_0$ was singled out on the basis of Yukawa coupling
unification in SO(10) GUT models.) Detection rates may
therefore be large in the focus point region.  This is especially true
for large $\tb$, where the effect noted in Sec.~\ref{sec:omega}, a
relatively light $H$ boson with $\tb$ enhanced $Hf\bar{f}$ couplings,
leads to large and dominant heavy Higgs amplitudes~\cite{K}.

\begin{figure}[t] 
\postscript{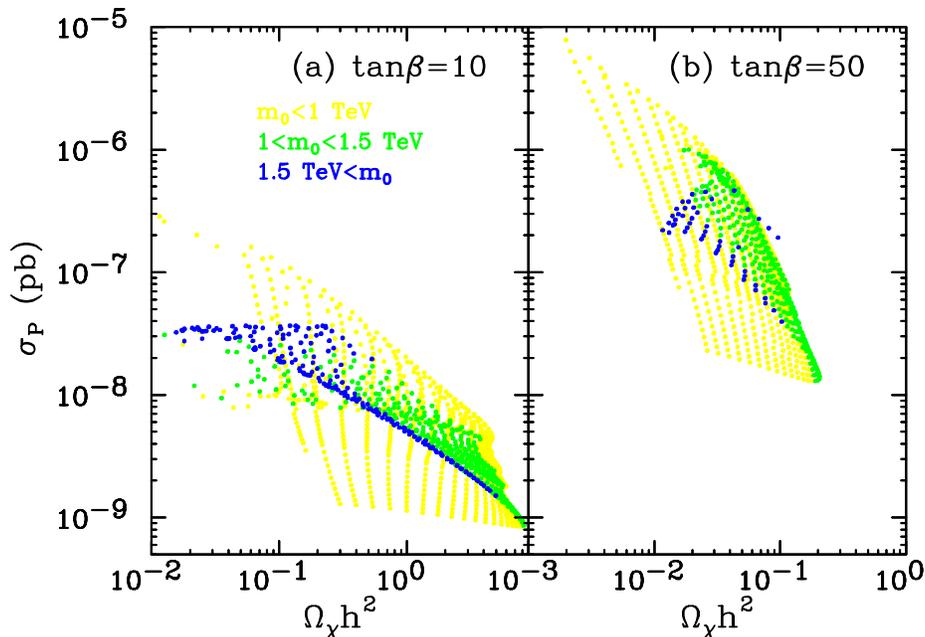}{0.74}
\caption{Regions of the $(\Omegachi h^2, \sigma_P)$ plane populated by
minimal supergravity models with $m_0 \leq 1\tev$ (yellow, light),
$1\tev < m_0 \leq 1.5\tev$ (green, medium), and $1.5\tev < m_0$ (blue,
dark). The parameters scanned are those given in
Fig.~\ref{fig:gfraction}, subject to the additional naturalness
constraint $M_{1/2} \leq 400\gev$.  We assume neutralino velocity
dispersion $\bar{v} = 270\ \text{km/s}$ and local density $\rho_0 =
0.3\ \text{GeV/cm$^3$}$, and $f_{T_s} = 0.14$ (see text).  }
\label{fig:omega_sigma}
\end{figure}

In Fig.~\ref{fig:omega_sigma}, we show the correlation between relic
density and proton-neutralino cross section for a representative
sample of points in parameter space.  As usual, there is a strong
anti-correlation, with large $\sigma_P$ corresponding to low
$\Omegachi h^2$.  Nevertheless, we find that, for a given $\tb$ and
$\Omegachi h^2$, points with large $m_0$ have proton-neutralino cross
sections comparable to those with conventional sub-TeV $m_0$.  For the
optimal case of large $\tb$, points with $m_0 > 1 \tev$ and $\Omegachi
h^2 > 0.1$ ($\Omegachi h^2 > 0.025$) have cross sections as large as
$\sigma_P \sim 2 \times 10^{-7}$ pb ($\sigma_P \sim 10^{-6}$ pb).

\section{Discussion}
\label{sec:discussion}

Several ongoing and planned experiments, using a variety of
techniques, are devoted to the search for WIMP dark
matter~\cite{Marina}.  Assuming that the matter-WIMP interactions are
dominated by spin-independent couplings, these experiments may be
compared by scaling the matter-WIMP cross sections for each detector
material to the proton-WIMP cross section $\sigma_P$ and displaying
the sensitivity in the $(m_{\chi}, \sigma_P)$ plane.

\begin{figure}[t]
\postscript{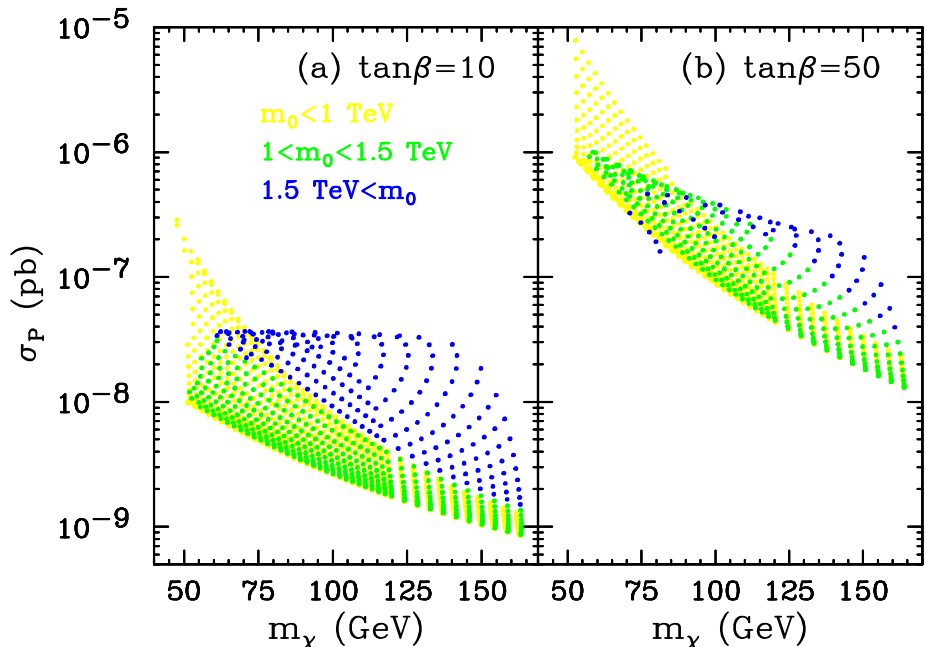}{0.74}
\caption{Points in minimal supergravity parameter space in the
$(m_{\chi}, \sigma_P)$ plane.  The parameters scanned, symbols, and
assumptions are as in Fig.~\ref{fig:omega_sigma}.}
\label{fig:mchi0_sigma}
\end{figure}

In Fig.~\ref{fig:mchi0_sigma} we plot predictions of focus point
supersymmetry in the $(m_{\chi}, \sigma_P)$ plane.  We find that, for
a given $m_{\chi}$, cross sections are typically maximized for large
$m_0$, and large cross sections near current sensitivities are
possible.  At present, the DAMA Collaboration has reported data
favoring the existence of a WIMP signal in their search for annual
modulation~\cite{DAMA}.  When WIMP velocity uncertainties are
included~\cite{BR}, the preferred range of parameters is $10^{-6}\
\text{pb} \alt \sigma_P \alt 10^{-5}\ \text{pb}$ and $30 \gev \alt
m_{\chi} \alt 200 \gev$ ($3\sigma$ CL).  The DAMA result has been
criticized~\cite{DAMAcomments}, and a recent analysis of the CDMS
Collaboration excludes the $3\sigma$ DAMA region at $> 84 \%$
CL~\cite{CDMS}.  While the situation is at present unclear, in the
near future both experiments will improve their sensitivities
significantly.  It is noteworthy, in any case, that current
experiments are on the verge of probing the region of parameter space
arising in focus point models.

\begin{figure}[tb]
\postscript{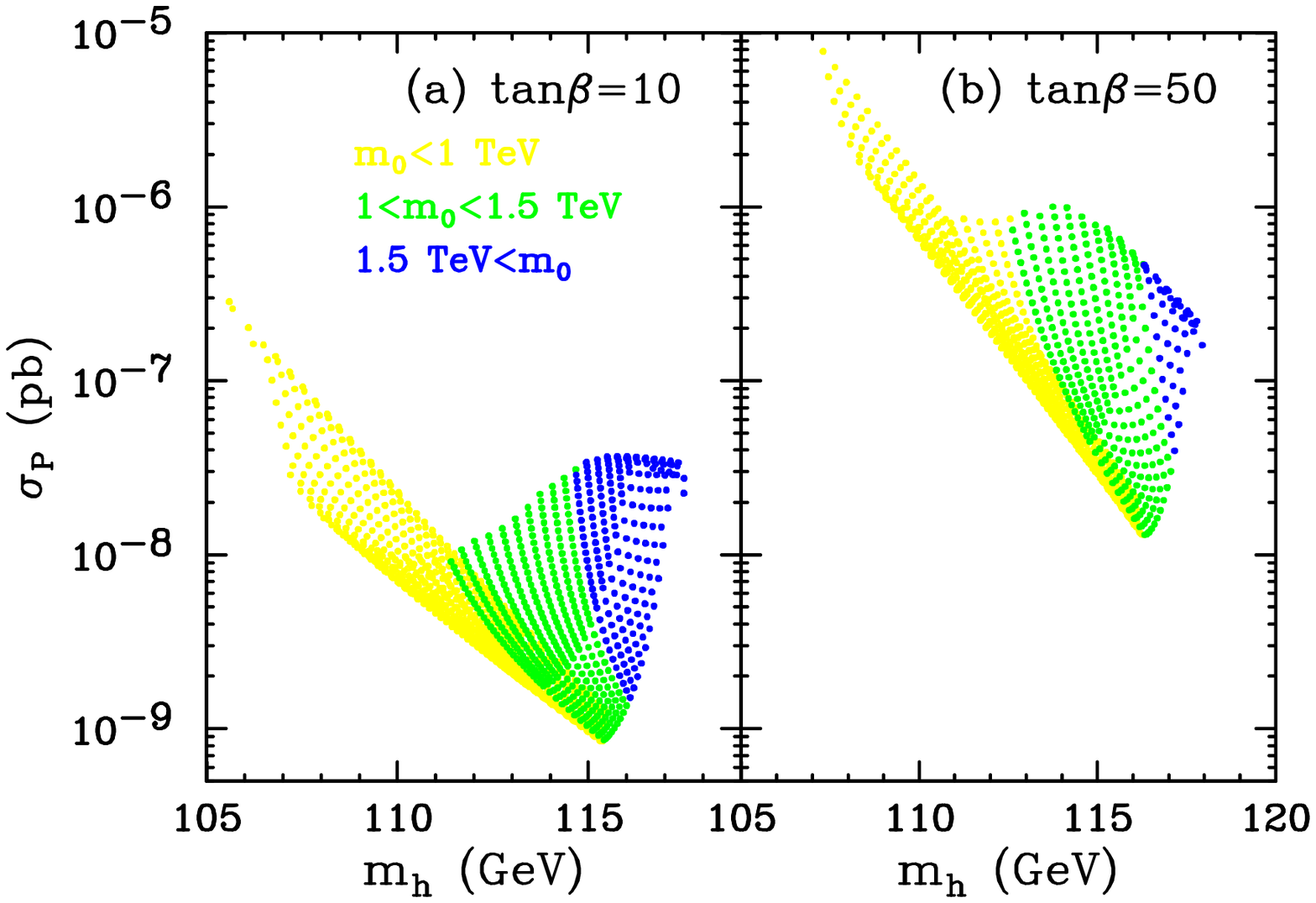}{0.74}
\caption{Points in minimal supergravity parameter space in the $(m_h,
\sigma_P)$ plane.  The parameters scanned, symbols, and assumptions
are as in Fig.~\ref{fig:omega_sigma}.}
\label{fig:mh_sigma}
\end{figure}

It is also interesting to compare the discovery potentials of dark
matter searches and future colliders in focus point models.  In the
focus point scenario, all squarks and sleptons are beyond the reach of
the Tevatron and may present significant challenges even for the LHC.
However, the lightest Higgs boson has mass $m_h \alt 120 \gev$, and so
may be discovered in Run II of the Tevatron. The correlation between
$m_h$ and $\sigma_P$ is presented in Fig.~\ref{fig:mh_sigma}.  We note
that focus point models provide a natural setting for Higgs masses
above the current bound, as the large squark masses raise the Higgs
mass through their radiative corrections. At the same time, both Higgs
and dark matter searches are promising, and will confront the focus
point scenario over the next few years.

It has been argued that the regions of parameter space which give
large dark matter detection rates may also predict large rates for
other processes, such as proton decay~\cite{protondecay} and $b\to s
\gamma$~\cite{bsg}.  Conversely, present bounds from such processes
may exclude the largest possible rates for dark matter detection.  The
proton decay rate, as noted in the introduction, scales as
$m_{\tilde{q}}^{-4}$, so it is relatively suppressed in the focus
point models.  The process $b \to s\gamma$ provides a stringent
constraint on large $\tb$ only for one sign of $\mu$.  For the
positive sign (the one adopted in our plots), destructive interference
between the chargino and charged Higgs diagrams typically leads to
predictions within the experimental bounds.

Focus point scenarios can also imply interesting rates for anti-matter
detectors and experiments sensitive to dark matter annihilation in the
cores of the earth or sun~\cite{Marina}.

Finally, of course, neutralinos and charginos themselves are targets
for discovery at colliders.  If they are found, detailed measurements
of their masses and compositions could then be made, and might even be
used to exclude neutralinos as dark matter candidates~\cite{collider}.
However, colliders will never be able to distinguish unstable
particles with long decay lengths from those stable on cosmological
time scales.  Since some well-motivated models with low energy
supersymmetry breaking predict LSPs which are stable on accelerator
detector, but unstable on cosmological, scales, in principle
identification of neutralinos as the dark matter of our universe
cannot be made at accelerators.  Ultimately this identification
requires detection, either directly or indirectly, of neutralinos
permeating space.

\section*{Acknowledgments}

We thank M.~Drees, C.~Kolda, and T.~Moroi for helpful conversations.
We are grateful to G.~Jungman for correspondence and M.~Drees and
T.~Falk for extensive numerical comparisons that revealed a sign error
in the program Neutdriver used in the first version of the paper.
This work was supported in part by the Department of Energy under
contracts DE--FG02--90ER40542 and DE--AC02--76CH03000, by the National
Science Foundation under grant PHY--9513835.  J.L.F. acknowledges the
generosity of Frank and Peggy Taplin and the National Center for
Theoretical Sciences and National Tsing Hua University, Taiwan for
hospitality during the completion of this work.  K.T.M. thanks the
SLAC theory group for hospitality during the completion of this work.

\end{document}